\documentclass[12pt]{iopart}
\usepackage{graphicx}
\begin{document}

\title[Critical exponent for the Anderson transition...]{Critical exponent for the Anderson transition in the three dimensional orthogonal universality class}

\author{Keith Slevin$^1$ and Tomi Ohtsuki$^2$}

\address{$^1$ Department of Physics, Graduate School of Science, Osaka University,
1-1 Machikaneyama, Toyonaka, Osaka 560-0043, Japan}
\address{$^2$ Department of Physics, Sophia University, Kioi-cho 7-1, Chiyoda-ku,
Tokyo 102-8554, Japan}
\ead{slevin@phys.sci.osaka-u.ac.jp}

\begin{abstract}
We report a careful finite size scaling study of the metal insulator transition in Anderson's model of localisation.
We focus on the estimation of the critical exponent $\nu$ that describes the divergence of the localisation length.
We verify the universality of this critical exponent for three different distributions of the random potential: box, normal and Cauchy.
Our results for the critical exponent are consistent with the measured values obtained in
experiments on the dynamical localisation transition in the quantum kicked rotor realised  in a cold atomic gas.
 \end{abstract}


\maketitle

\section{Introduction}

The zero temperature dc conductivity $\sigma$ of a solid can be expressed using the Einstein relation
\begin{equation}
\label{eq:einsteinrelation}
  \sigma = e^2 \rho  \left( E_\mathrm{F} \right) D \;.
\end{equation}
Here $\rho \left( E_\mathrm{F} \right)$ is the density of states per unit volume at the Fermi level and $D$ is the diffusion constant.
The electrical conductivity of weakly disordered materials can be understood using a semiclassical picture.
The motion of electrons on the scale of the lattice constant is quantum mechanical,
while the motion on the scale of the (much longer) mean free path is classical.
For classical diffusion, the diffusion constant is equal to
\begin{equation}\label{eq:diffusionconstant}
  D = \frac{1}{d} v_\mathrm{F} \ell \;,
\end{equation}
where $d$ is the dimensionality, $v_\mathrm{F}$ the Fermi velocity, and $\ell$ is the mean free path.
Provided the diffusion constant is not zero, we expect the material to be a metal unless
the Fermi level lies in a band gap.
Then, dependent on the size of the gap, we expect the material to be either an insulator or a semiconductor.
The occurrence of a band gap may be explainable within a single particle picture,
or it may be the result of correlation effects \cite{vladbook}.

Is it possible that the diffusion constant becomes zero? For classical diffusion, the answer is no.
For strong disorder the mean free path can be short but it is always finite and so is the diffusion constant.
Anderson \cite{anderson58} was the first to realise that for quantum diffusion the situation is different.
In this case, quantum interference may result in the complete suppression of diffusion
even though the mean free path remains finite.
This effect is now called Anderson localisation \cite{abrahams10}.
The suppression of diffusion is a reflection of a change in the nature of the electronic eigenfunctions.
The effect is particularly pronounced in lower dimensions.
For one and two-dimensional systems the eigenfunctions are,
apart from some special cases, always exponentially localised in space.
In three dimensions, eigenfunctions are localised only for sufficiently strong disorder.
There is thus a metal insulator transition as the strength of the disorder is increased.
This transition is called the Anderson transition and it is an example of a zero temperature continuous quantum phase transition.

In common with continuous thermal phase transitions, the concept of universality class plays a central role.
In the vicinity of the critical point various critical phenomena described by power laws occur.
The exponents appearing in these power laws are expected to be universal, i.e. to depend only on the universality class.
The universality classes are determined by the dimensionality of the system and the symmetries of the Hamiltonian.
Since we are considering disordered systems, the Hamiltonian does not have any translational symmetry.
Rather, the important symmetries for the Anderson localisation problem are time reversal symmetry and spin rotation symmetry.
In addition, certain discrete symmetries may also play a role \cite{gade91,gade93}.
If these discrete symmetries are ignored, we arrive at the three Wigner-Dyson symmetry classes: orthogonal, unitary and symplectic.
If discrete symmetries are included, ten symmetry classes need to be considered \cite{zirnbauer96,altland97}.
%Further details are given in the Appendix.
In this paper, we report a finite size scaling study of
Anderson's model of localisation.
This is defined on a three dimensional lattice and the Hamiltonian has both time reversal and spin-rotation symmetries.
Thus, our focus is on the three dimensional orthogonal universality class.

The critical phenomena of the Anderson transition are described by two independent critical exponents.
The first of these is the critical exponent $\nu$ that describes the divergence of the correlation length at the transition
\begin{equation}
\label{eq:xiDivergence}
\xi \sim \frac{1}{|x-x_\mathrm{c}|^\nu}\;.
\end{equation}
Here, $x$ is the parameter that is varied to drive the transition, and $x_\mathrm{c}$ is its critical value.
The second is the dynamic exponent $z$ that describes how frequency is re-normalised near the critical point.
For models where electron-electron interactions are neglected,
the only relevant energy scale at the transition is the level spacing.
From this it follows that the dynamic exponent is equal to the dimensionality $z=d$.
While concerted efforts have been made to calculate the critical exponent $\nu$ using an $\epsilon$
expansion about the lower critical dimension $(d=2)$ \cite{hikami81,bernreuther86,wegner89},
reliable values have not been obtained in this way.
This gap has been filled by extensive numerical simulations.

In addition to the critical phenomena mentioned above, scaling of the wavefunction intensity distribution with
system size at the transition is described by a multi-fractal spectrum \cite{evers08}.
This multi-fractal spectrum, which is again expected to display universality, has also been the object of careful numerical study \cite{alberto09,alberto10,alberto11}.

This article is concerned with the Anderson localisation of electrons. However, Anderson localisation is a wave phenomenon and is also observable for classical waves \cite{shengbook}.
The periodically driven quantum kicked rotor exhibits an analogue of localisation in momentum space called dynamic localisation \cite{haakebook}.
Moreover, if the amplitude of the kick is modulated quasi-periodically in an appropriate way
\cite{Casati89}, it is possible to observe a dynamical localisation transition which is believed to be in the same universality class as the Anderson transition in the model we study here \cite{chabe08,lopez12}.

\section{Model and Method}
\subsection{Anderson's model of localisation}

The Hamiltonian for Anderson's model of localisation is \cite{anderson58}
\begin{equation}\label{eq:hamiltonian}
  H = \sum_{i} W_{i} c_{i}^{\dagger} c_{i} - V \sum_{\left<ij\right>} c_{i}^{\dagger} c_{j}\;.
\end{equation}
Here, the sum in the first term is over the sites of a three dimensional
 simple cubic lattice with lattice constant $a$. The sum in the second term is over nearest neighbour lattice sites.
The creation operator $c_i^{\dagger}$ creates an electron in an orbital $\left| i \right>$ that is localised on-site $i$. Orbitals on different sites are assumed to be orthogonal. The state vector of the system is then
\begin{equation}\label{eq:statevector}
  \left| \psi \right> = \sum_{i} \psi_{i} \left| i \right>\;.
\end{equation}
If the boundary conditions are specified, the eigenstates and eigenenergies may be found by solving the time-independent Schr\"{o}dinger equation
\begin{equation}\label{eq:schrodinger}
  H \left| \psi \right> = E \left| \psi \right>\;.
\end{equation}
The constant $V$ sets the energy scale.
We take $a$ as the unit of length, so that
\begin{equation}\label{eq:unitlength}
  a=1\;,
\end{equation}
 and $V$ as the unit of energy, so that
\begin{equation}\label{eq:unitenergy}
  V=1\;.
\end{equation}
The on-site potentials $W_i$ are independently and identically distributed random variables with distribution
\begin{equation}\label{eq:probabilitydistribution}
  P\left( W_i \right) = p \left( W_i \right) d W_{i}\;.
\end{equation}
We consider three distributions.
The first is the box distribution
\begin{equation}\label{eq:boxdistribution}
  p \left( W_{i} \right) = \left\{
  \begin{array}{cc}
    1/W & \left| W_{i} \right| \leq W/2 \\
    0 & \mathrm{otherwise}
  \end{array}
  \right.\;.
\end{equation}
The parameter $W$ characterises the strength of the disorder. In what follows, we usually refer to $W$ simply as the disorder.
The second is the normal distribution
\begin{equation}\label{eq:normaldistribution}
  p \left( W_{i} \right) = \frac{1}{\sqrt{2\pi \sigma^2 }} \exp\left( -\frac{W_i^2 }{ 2 \sigma^2} \right)\;.
\end{equation}
To make it easier to compare with the box distribution, we set the variance of the normal distribution equal to that of the box distribution
\begin{equation}\label{eq:normalvariance}
\sigma^2 = \frac{W^2}{12}\;.
\end{equation}
The third is the Cauchy distribution
\begin{equation}\label{eq:cauchydistribution}
  p \left( W_{i} \right) = \frac{W}{\pi \left( W_i^2 + W^2 \right) }\;.
\end{equation}
The parameter $W$ again specifies the strength of the disorder. However, since the Cauchy distribution does not have a second moment, its value is not directly comparable with that of the box and normal distributions.

\subsection{The transfer matrix method}
Rather than studying the eigenstates of the Anderson model directly,
we consider the transmission of electrons with a given energy $E$ through long disordered wires
with a uniform square cross-section
\begin{equation}
  L_y = L_z = L\;.
\end{equation}
As a consequence of Anderson localisation, for wires that are sufficiently long,
the amplitude of the transmission decays exponentially with an
associated decay length called the quasi-one-dimensional localisation length $\xi_\mathrm{Q1D}$.
For a given distribution,  $\xi_\mathrm{Q1D}$ is a function of the energy, disorder and the
transverse dimension $L$
\begin{equation}
 \xi_\mathrm{Q1D} \equiv \xi_\mathrm{Q1D}(E,W,L) \;.
\end{equation}
We expect the exponential decay of the transmission to be observable when
\begin{equation}
  L_x \gg \xi_\mathrm{Q1D}\;.
\end{equation}
We use the transfer matrix method to estimate $\xi_\mathrm{Q1D}$ \cite{mackinnon81,pichard81,mackinnon83}.

We divide the system into slices labelled by their position $x$ in the $x$-direction.
Since the off-diagonal elements of the Hamiltonian are nonzero only between nearest neighbour sites,
the time-independent Schr\"odinger equation reduces to a set of linear equations relating the wave function amplitudes on adjacent slices.
These equations can be rewritten in the form of a transfer matrix multiplication
\begin{equation}\label{eq:transfervector}
  \left[
  \begin{array}{c}
    \psi_{x+1} \\
    -\psi_{x}
  \end{array}
  \right]
  = M_x
  \left[
  \begin{array}{c}
    \psi_x \\
    -\psi_{x-1}
  \end{array}
  \right] \;.
\end{equation}
Here, $\psi_x$ groups together all the wavefunction amplitudes $\psi_i$ on the cross section through the wire at position $x$
\begin{equation}\label{eq:psix}
  \psi_x = \left[
  \begin{array}{c}
    \vdots \\
    \psi_i \\
    \vdots
  \end{array}
  \right] \;,
\end{equation}
and $M_x$ is the transfer matrix defined by
\begin{equation}\label{eq:transfermatrix}
  M_x = \left[
  \begin{array}{cc}
    \left< x \right| H \left| x \right> - E & 1 \\
    -1 & 0
  \end{array}
  \right] \;.
\end{equation}
Since there are
\begin{equation}\label{}
  L_y \times L_z = L^2 \equiv N
\end{equation}
wavefunction aplitudes on each slice,
the size of the transfer matrix is $2N\times 2N$.
The transfer matrix for a given layer $x$ is a function of the energy $E$ and the on-site potentials $W_i$ of
lattice sites on the slice.
The boundary conditions in the $y$ and $z$ directions also need to be specified.
Throughout this work, we impose periodic boundary conditions in these directions.

To estimate the quasi-one-dimensional localisation length
we consider the Lyapunov exponents of the following random matrix product
\begin{equation}\label{eq:transfermatrixproduct}
  M = \prod_{x=1}^{L_x} M_x \;.
\end{equation}
From this product we define a real symmetric matrix
\begin{equation}\label{eq:omega}
  \Omega = \ln MM^T \;.
\end{equation}
As a consequence of current conservation the eigenvalues of this matrix occur in pairs of opposite sign.
In what follows, we shall suppose that the eigenvalues $\nu_i$ ($i=1,\cdots,2N$) are numbered in decreasing order,
i.e. so that $\nu_i > \nu_j$ when $i < j$.
For this ordering, current conservation means that
\begin{equation}\label{eq:omegaeigenvalues}
  \nu_{N+i} = -\nu_{N-i+1}
\end{equation}
with $i=1,\cdots, N$.
The Lyapunov exponents associated with the random matrix product are defined by taking the following limit
\begin{equation}\label{eq:lyapunovs}
  \gamma_i = \lim_{L_x \rightarrow \infty} \frac{\nu_i}{2L_x} \;.
\end{equation}
In numerical simulations, for practical reasons (see below), the Lyapunov exponents are not calculated by diagonalising the matrix $\Omega$.
Instead, to estimate the first $m$ largest Lyapunov exponents we start with a $2N \times m$ matrix $U$ with orthogonal columns and consider the QR decomposition of the matrix
\begin{equation}\label{eq:qrfactorization}
  MU=QR\;.
\end{equation}
Here, $Q$ is a $2N \times m$ matrix with orthogonal columns and $R$ is a $m \times m$ upper triangular matrix with positive elements on the diagonal.
The Lyapunov exponents are related to the diagonal elements of $R$ by
\begin{equation}\label{eq:lyapunovsfromR}
  \gamma_i = \lim_{L_x \rightarrow \infty} \frac{1}{L_x} \ln R_{i,i}\;.
\end{equation}
In practice, we estimate the Lyapunov exponents by truncating the transfer matrix multiplication at a large but finite length $L_x$
\begin{equation}\label{eq:lyapunovestimate}
  \tilde{\gamma}_i = \frac{1}{L_x} \ln R_{i,i}\;.
\end{equation}
Here, the tilde denotes that this is an estimate.
Note that these estimates do not necessarily obey the exact symmetry of equation (\ref{eq:omegaeigenvalues}).
In general, this is recovered only in the limit $L_x \rightarrow\infty$.
Nevertheless, if we estimate all the Lyapunov exponents by setting $m=2N$, we find that the sum of the exponents is exactly zero for any $L_x$
\begin{equation}\label{eq:sumestimates}
  \sum_{i=1}^{2N} \tilde{\gamma}_i = 0 \;.
\end{equation}
This is proven by taking the determinant of equation (\ref{eq:qrfactorization}).
It is also helpful to randomise the starting vectors $U$ by performing several transfer matrix multiplications and QR factorisations, and then replacing $U$ with the final $Q$ matrix.
For a more complete discussion of these technicalities see \cite{slevin04}.

To relate the Lyapunov exponents with the quasi-one dimensional localisation length, we consider the following scattering problem. Suppose that the wire is connected to perfect leads at both left and right.
If we consider an outgoing state at, say, the left of the sample, we can use the transfer matrix multiplication to calculate the wavefunction amplitudes in the right lead.
This observation is the basis for a practical method for the calculation of the transmission and reflection matrices for this problem \cite{pendry92}.
It also makes clear that the decay of the transmission amplitude will be related to the slowest decay rate in the problem, i.e.,
the smallest positive Lyapunov exponent.
It is thus usual to identify the reciprocal of the smallest positive Lyapunov exponent with the quasi-one dimensional localisation length
\begin{equation}\label{eq:sple}
  \xi _{\rm{Q1D}} = \frac{1}{\gamma _N} \;.
\end{equation}
Let us just note in passing that, while when performing a finite size scaling analysis (as will be described below) it is customary to consider only the smallest positive Lyapunov exponent, similar analyses of the second, third, etc. smallest positive exponent lead to the same results \cite{slevin01a}.

\subsection{Dealing with round-off error}
Unfortunately the transfer matrix multiplication is unstable and, as a result, is very sensitive to round-off error.
This means that direct calculation of the Lyapunov exponents by diagonalising the matrix $\Omega$, or QR
factorisation of the transfer matrix product $M$, is not possible.
The standard way to overcome this difficulty is to perform repeated QR factorisations at intervals throughout the transfer matrix multiplication.
If the the number of transfer matrix multiplications between each QR factorisation is sufficiently small, the loss of precision due to round-off error can be avoided.
We start the transfer matrix multiplication with
\begin{equation}\label{eq:qzero}
  Q^{\left( 0 \right)} = U
\end{equation}
and perform QR factorisations after every $q$ iterations
\begin{equation}\label{eq:qriteration}
  Q^{\left( j \right)} R^{\left( j \right)} = M_{jq} \cdots M_{\left(j-1\right)q + 1} Q^{\left( j-1 \right)} \; \;
  \left( j = 1, \cdots, l \right) \;.
\end{equation}
For simplicity we assume here that $L_x$ is an integer multiple $l$ of $q$, i.e. that
\begin{equation}\label{eq:mdef}
  L_x = lq \;.
\end{equation}
The estimates of the Lyapunov exponents are then
\begin{equation}\label{eq:raverage}
  \tilde{\gamma}_i = \frac{1}{L_x} \sum_{j=1}^{l} \ln R_{i,i}^{\left( j \right)}\;.
\end{equation}
Provided $q$ is small enough to control round-off error, the estimates of the Lyapunov exponents are independent of $q$.

In an attempt to determine a reasonable value for $q$ systematically, we checked how large we can make $q$ while still maintaining the sum of the
Lyapunov exponents close to zero. This is illustrated in Tables \ref{table:boxq} and \ref{table:cauchyq} for the box and Cauchy distributions. For the Cauchy distribution, in particular, the effect of round-off error is clearly visible in the sum  unless $q$ is very small.
We also show the estimates of the smallest positive Lyapunov exponent.
It is noticeable that the estimates of the smallest positive exponent are much less sensitive to round-off error than
the sum of all the exponents.
Nevertheless, we preferred to be cautious and so chose $q$ to ensure that the sum of exponents is zero to about single precision accuracy.

\begin{table}[htb]
\begin{center}
\begin{tabular}{|l|l|l|}
  \hline
  $q$ & $\Sigma \tilde{\gamma}_i$ & $\tilde{\gamma}_N$ \\ \hline
  3 & $10^{-13}$    & 0.10380011865 \\
  6 & $10^{-9}$     & 0.10380011866 \\
  9 & $10^{-4}$     & 0.10380011848 \\
  \hline
\end{tabular}
\caption{An example of the determination of the interval $q$ at which to perform the QR factorisation in order to control round-off error. The data are for the box distribution with $L_x=9000$, $L_y=L_z=24$, $W=18$ and $E=1$.
The starting vectors $U$ were randomized with 100 transfer matrix multiplications.}
\label{table:boxq}
\end{center}
\end{table}

\begin{table}[htb]
\begin{center}
\begin{tabular}{|l|l|l|}
  \hline
  $q$ & $\Sigma \tilde{\gamma}_i$ & $\tilde{\gamma}_N$ \\ \hline
  1 & $10^{-13}$ & 0.0953267557872 \\
  2 & $10^{-7}$ & 0.0953267557874 \\
  4 & $10^{-3}$ & 0.0953267558688 \\
  \hline
\end{tabular}
\caption{An example showing that it is more difficult to control round-off error for the Cauchy distribution.
Here, $L_x=1000$, $L_y=L_z=24$, $W=4.5$ and $E=0$.
The starting vectors $U$ were randomized with 100 transfer matrix multiplications.}
\label{table:cauchyq}
\end{center}
\end{table}

\subsection{Determination of the precision of the Lyapunov exponents}

To perform the finite size scaling analysis, estimates of the Lyapunov exponents alone are insufficient.
In addition, accurate estimates of the precision of the estimates of the Lyapunov exponents are required.
In most previous work these estimates were obtained by supposing that the terms appearing in the sum
(\ref{eq:raverage}) are statistically independent.
This assumption is not unreasonable if the interval $q$ between QR factorisations is sufficiently large.
However, this is not always the case.
Particularly for the Cauchy distribution it is necessary to set $q$ to a small value in order to avoid round-off error.
In this case, the assumption of statistical independence leads to erroneously small error estimates.
This is serious because it means that reliable estimation of the quality of the finite size scaling fit using the goodness of fit probability is not possible.
To circumvent this difficulty we define
\begin{equation}\label{eq:dk}
  D_i^{\left( k \right)} = \sum_{j=\left( k - 1 \right)r+1}^{kr} \frac{1}{p} \ln R_{i,i}^{\left( j \right)}
   \; \;
  \left( k = 1, \cdots, s \right)
\end{equation}
where $r$ is an integer, $p=qr$, and we have supposed for simplicity that $L_x=ps$ with $s$ an integer.
The estimates of the Lyapunov exponents are then given by the mean values of the $D_i$,
\begin{equation}\label{eq:daverage}
  \tilde{\gamma}_i = \frac{1}{s} \sum_{k=1}^{s} D_i^{\left( k \right)} \equiv \overline{D} \;.
\end{equation}
If $p$ is sufficiently large, the assumption that the $D_i^{\left( k \right)}$ are statistically independent for different $k$ is a reasonable approximation.
The precision of the estimates of the Lyapunov exponents are then given by the usual formulae for the standard error $\sigma_i$ of the mean
\begin{equation}\label{eq:error}
  \sigma_i^2 = \frac{1}{s-1} \left( \overline{D^2} -  \overline{D}^2 \right) \,
\end{equation}
where
\begin{equation}\label{eq:dsquared}
  \overline{D^2} \equiv \frac{1}{s} \sum_{k=1}^{s} \left( D_i^{\left( k \right)} \right)^2 \;.
\end{equation}
When performing the simulation, we decide in advance the required precision and stop the simulation
when the standard error given by equation (\ref{eq:error}) satisfies this criterion.
(We also take care that a sufficient number of transfer matrix multiplications are always performed to ensure that
we have sufficient statistics to estimate the standard error reliably using equation (\ref{eq:error}), i.e, we set a
minimum value for $s$.)

\begin{figure}
  \centering
  \includegraphics[width=0.8\textwidth]{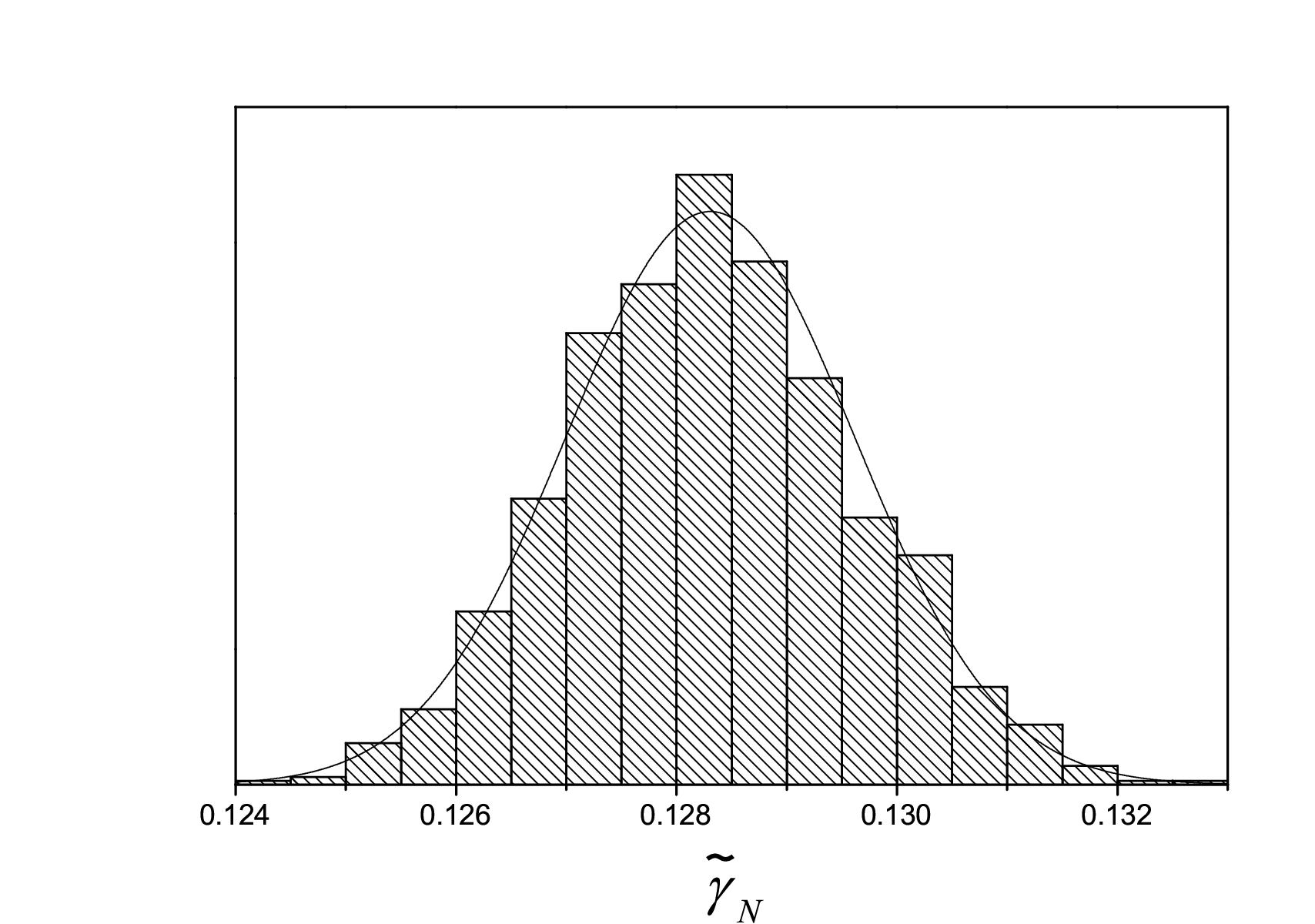}
  \caption{The distribution of the estimate of the smallest positive Lyapunov exponent obtained in repeated simulations with the same parameters with
  independent stream of random numbers. The results shown are for the box distribution with $L_y=L_z=10$, $W=15.0$ and $E=1$.
  QR factorizations were performed every $q=6$ transfer matrix multiplications and we set $r=5$ for the determination of the precision.}
\label{fig:errorhist}
\end{figure}

In Figure \ref{fig:errorhist}, we show some typical results. The simulation has been set to terminate when according to
equation (\ref{eq:error}) the precision is better than $1\%$.
The simulation has been repeated 100 times using independent streams of random numbers.
The observed fluctuation is $1.1\%$, which is within approximately a single standard deviation of the
expected value $1\%$.
Moreover, as expected from the form of equation (\ref{eq:daverage}), the normal distribution gives a reasonable description of the sample to sample fluctuations.

\subsection{Finite size scaling}

In our simulations we fix the energy $E$.
There is then a critical disorder $W_\mathrm{c}\equiv W_\mathrm{c}(E)$ that separates the localised and extended phases.
We then accumulate data for the smallest positive Lyapunov exponent for a range of disorder around the critical disorder
and for as wide a range of system sizes as practicable.
The finite size scaling method is then used to extract information about the critical phenomena from this numerical data.
This allows us to estimate universal properties such as the critical exponent, the value $\Gamma_\mathrm{c}$ of $\Gamma$ (to be explained below) at the critical point, the scaling function etc., as well as non-universal properties such as the critical disorder.
The starting point is to assume that the disorder and system sizes dependence of the dimensionless quantity
\begin{equation}\label{eq:gamma}
  \Gamma = \gamma_N L \;,
\end{equation}
are described by a scaling law of the form \cite{slevin99}
\begin{equation}\label{eq:fssfit}
  \Gamma = F \left( \phi_1, \phi_2 \right) \;.
\end{equation}
Here, $F$ is a universal function, the arguments $\phi_1$ and $\phi_2$ are scaling variables
\begin{equation}\label{eq:scalingvariables}
  \phi_i = u_i \left( w \right) L^{\alpha_i} \;,
\end{equation}
and $w$ is  the reduced disorder
\begin{equation}\label{eq:reduceddisorder}
  w = \frac{W - {W_\mathrm{c}}}{W_\mathrm{c}} \;.
\end{equation}
The first of the scaling variables is the relevant scaling variable and has associated with it a positive exponent $\alpha_1>0$.
The critical exponent is given by the inverse of this exponent
\begin{equation}\label{eq:criticalexponent}
 \nu = \frac{1}{\alpha_1} \;.
\end{equation}
The second is an irrelevant scaling variable and has associated with it a negative exponent $\alpha_2 < 0$.
The value of this exponent is usually denoted by the letter $y$
\begin{equation}\label{eq:irrelevantexponent}
   y \equiv \alpha_2 \;.
\end{equation}
The functions $u_i$ appearing in the scaling variables are expanded as Taylor series
\begin{equation}\label{eq:scalingvariablexpansion}
  {u_i}\left( w \right) = \sum\limits_{j = 0}^{{m_i}} {{b_{i,j}}{w^j}} \;.
\end{equation}
This allows us to take account of possible nonlinearity of the scaling variables in the disorder.
For the relevant scaling variable we must have
\begin{equation}\label{eq:u1}
  u_1\left( {w = 0} \right) = 0 \;.
\end{equation}
The scaling function is expanded as a Taylor series in the scaling variables
\begin{equation}\label{eq:scalingfunctionexpansion}
  F = \sum\limits_{{j_1} = 0}^{{n_1}} { \sum\limits_{{j_2} = 0}^{{n_2}} { {a_{{j_1}, {j_2} }}\phi _1^{{j_1}}  \phi _2^{{j_2}}  } }\;.
\end{equation}
To avoid ambiguity in the definition of the fitting model we set
\begin{equation}\label{eq:firstordercoefficients}
  a_{1,0} = a_{0,1} = 1 \;.
\end{equation}
In principle, there are many irrelevant scaling variables. However, in practice,
it is extremely difficult to resolve contributions of different irrelevant variables.
Therefore, we have made the approximation that the contribution of a single irrelevant variable dominates and neglected the others.

For the purposes of fitting data near the critical point we expect that it is reasonable to truncate the Taylor series at fairly low order.
The total number of parameters in the model is
\begin{equation}\label{eq:NP}
  N_{\mathrm{P}} = 2 + m_1 + m_2 + (n_1 + 1) (n_2 + 1 ).
\end{equation}
If the irrelevant variable is neglected in the analysis, the total number of parameters in the model becomes
\begin{equation}
  N_{\mathrm{P}}  = 2 + m_1 + n_1.
\end{equation}
To find the best fit, we start with reasonable initial estimates for the fitting parameters and use the
Levenberg-Marquardt algorithm to minimize the $\chi^2$ statistic
\begin{equation}\label{eq:chisqq}
  \chi^2 = \sum_{i=1}^{N_D} \frac{\left( F_i - \Gamma_i \right)^2}{\sigma_i^2} \;.
\end{equation}
Here $F_i$ is the value of finite size scaling model evaluated at the parameters used in the $i$th run of the simulation,
$\Gamma_i$ is the value of $\Gamma$ found in that simulation and $N_{\mathrm{D}}$ is the total number of simulations performed, i.e. the number of data.

The use of the $\chi^2$ statistic is rigorously justified provided the
deviations between the model and the data are independent normally distributed random errors,
and the model is linear in all the parameters.
In this case, the minimum value $\chi^2_{\mathrm{min}}$ of the  $\chi^2$ statistic  is distributed according to the $\chi^2$ distribution with
\begin{equation}\label{eq:dof}
  N_{\mathrm{DOF}} = N_{\mathrm{D}} - N_{\mathrm{P}}
\end{equation}
degrees of freedom.
The goodness of fit $P$, which is the probability that a worse value of $\chi^2_{\mathrm{min}}$
would be obtained by chance, is given by the formula (see Chapter 11 of the book by Bevington and Robinson \cite{Bevington})
\begin{equation}\label{eq:gof}
  P = \int_{\chi^2_{\mathrm{min}}}^{\infty} p_{\chi} \left( \chi^2, N_{\mathrm{DOF}} \right) d\chi^2
\end{equation}
where $p_{\chi}$ is the $\chi^2$ distribution with $N_{\mathrm{DOF}}$ degrees of freedom
\begin{equation}\label{eq:chisqdist}
   p_{\chi} \left( \chi^2, N_{\mathrm{DOF}} \right) = \left( \chi^2 \right)^{\left( N_{\mathrm{DOF}}-2 \right)/2  } \frac{ \exp \left( -\chi^2/2 \right) }{ 2^{N_{\mathrm{DOF}}/2} \Gamma \left( N_{\mathrm{DOF}}/2 \right) } \;.
\end{equation}
This is calculated as described in \cite{NRfortran}; we have
\begin{equation}
  P = 1 - P \left( N_{\mathrm{DOF}}/2, \chi_\mathrm{min}^2/2 \right)
\end{equation}
where
\begin{equation}
  P\left( a, x \right) = \frac{1}{\Gamma\left( a \right) }{ \int_{0}^{x} \exp \left( -t \right)  t^{a-1} d t}
\end{equation}
is the incomplete Gamma function and $\Gamma\left(a\right)$ is the Gamma function.

Here, in fact, the model is linear in some but not all of the parameters.
We proceed on the basis that the data are sufficient to narrow down the possible values of the fitting parameters to
a small enough region such that, at least in principle, the model could be replaced by a linear approximation.
This is true for the most important parameters such as the critical exponent, disorder etc., but not for all the
parameters.
To determine the precision of the fitted parameters we use the Monte Carlo method described in \cite{NRfortran}.
This involves the generation and fitting of large numbers of pseudo-data sets.
We give the errors in the form of $95\%$ confidence intervals.\footnote{In this article, errors expressed using square brackets are
95\% confidence intervals, while numbers appearing after the symbol $\pm$ are standard errors.}
This method also allows us to check the goodness of fit probability by calculating it directly
from the histogram of $\chi^2$ obtained in the Monte Carlo simulation.
We have found no significant difference with the values obtained from equation (\ref{eq:gof}).

\section{Results of the finite size scaling analysis}

For the three distributions of the random potential, we simulated systems with sizes $L\times L \times L_x$ with
$L=4, 6, 8, 10, 12, 16, 20, 24$.
In each case the simulations were terminated when Eqs. (\ref{eq:daverage}) and (\ref{eq:error}) indicated that the
precision of the estimate of the smallest positive Lyapunov exponent had reached $0.1\%$.
This was typically of the order of $L_x = 10^6 \sim 10^7$ transfer matrix multiplications.
The energies, disorder ranges and total numbers of data points are listed in Table \ref{table:details}.
For the box and normal distributions of the random potential, QR factorizations were performed every $q=6$ transfer matrix multiplications.
The precision of the estimate of the smallest positive Lyapunov exponent was checked after every $r=5$ factorizations,
i.e. after every $p=30$ transfer matrix multiplications.
For the Cauchy distribution of the random potential, QR factorizations were performed every $q=2$ transfer matrix multiplications and
the precision was estimated after every $r=10$ factorizations, i.e. after every $p=20$ transfer matrix multiplications.

The data for each distribution of the random potential were then fitted using the finite size scaling model described above.
For the box distribution of the random potential, the starting values used in the non-linear least squares fitting were
$W_\mathrm{c}=16.53$, $\Gamma_\mathrm{c}=1.73$,  $\alpha_1=0.63$, $b_{1,1}=1.3$, $\alpha_2=-2.5$ and $b_{2,0}=-0.3$.
For the normal distribution of the random potential, the starting values were
$W_\mathrm{c}=6.15$, $\Gamma_\mathrm{c}=1.73$,  $\alpha_1=0.63$,  $b_{1,1}=1.1$, $\alpha_2=-1.5$ and $b_{2,0}=-1.0$.
And for the Cauchy distribution, the starting values were $W_\mathrm{c}=4.3$, $\Gamma_\mathrm{c}=1.72$,
$\alpha_1=0.63$,  $b_{1,1}=1.0$, $\alpha_2=-2.5$ and $b_{2,0}=0.2$.
All the other parameters were initially set to zero.

\begin{figure}
  \centering
  \includegraphics[width=0.8\textwidth]{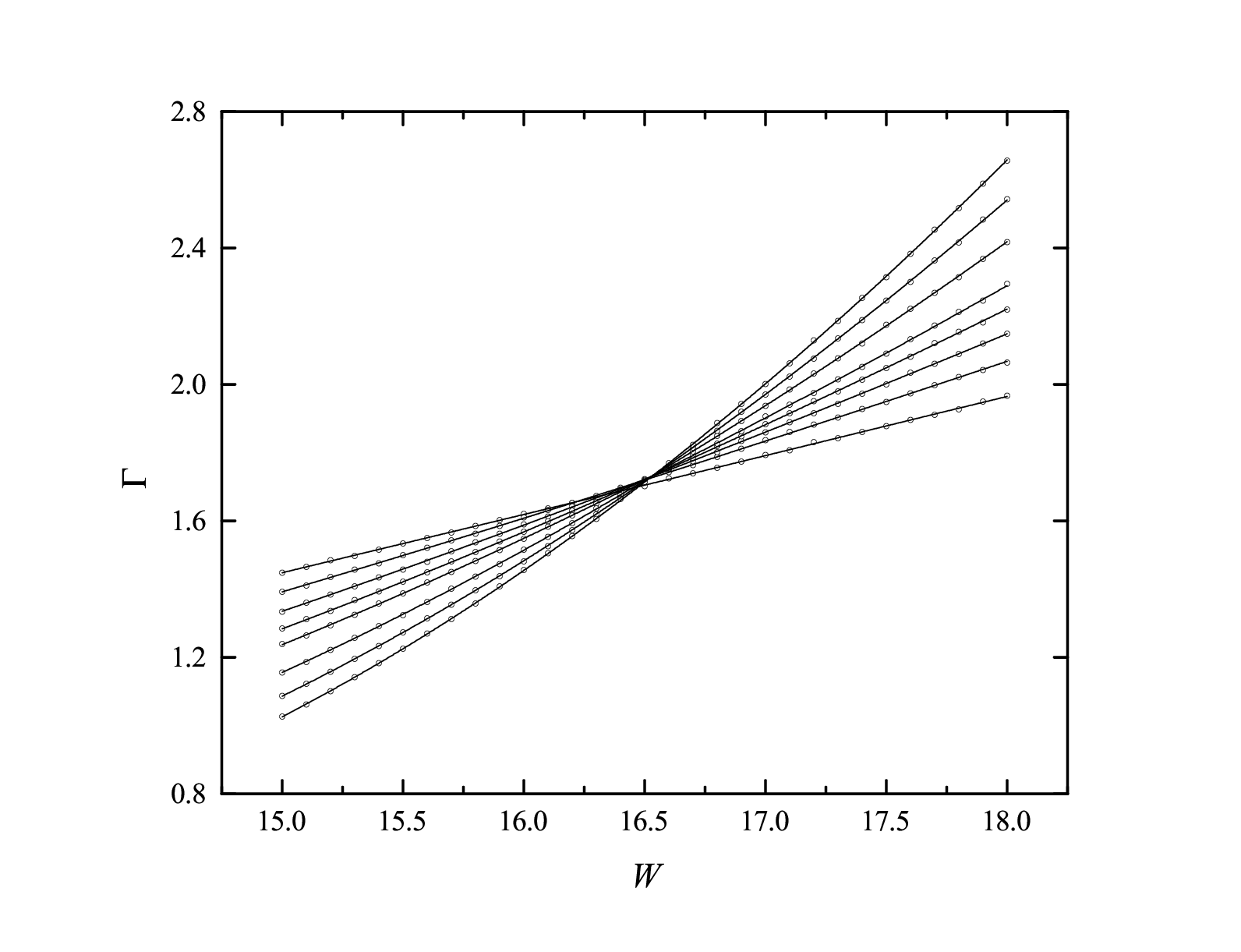}\\
  \caption{Fit of the data for the box distribution of the random potential.}\label{fig:abx}
\end{figure}

\begin{figure}
  \centering
  \includegraphics[width=0.8\textwidth]{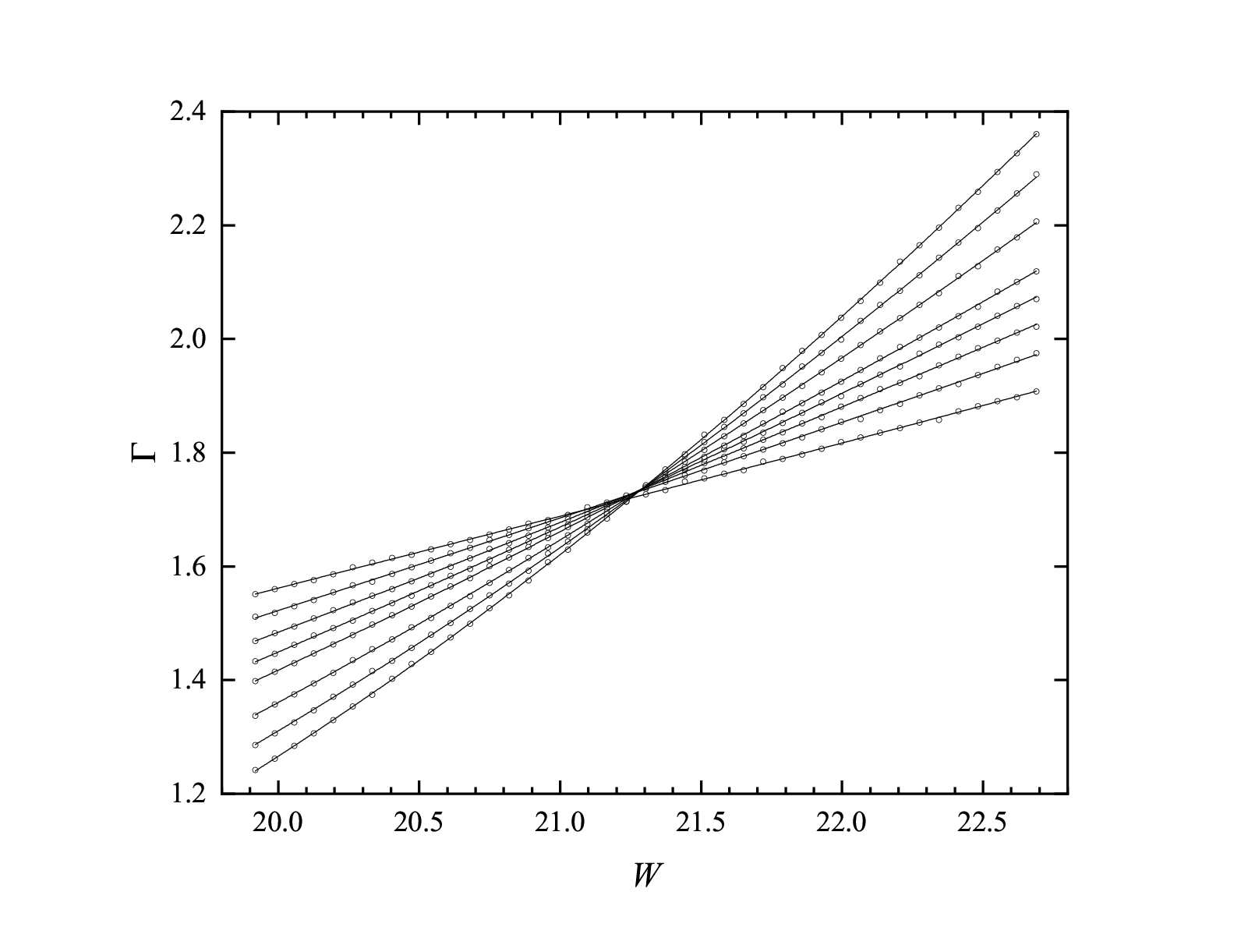}\\
  \caption{Fit of the data for the normal distribution of the random potential.}\label{fig:anm}
\end{figure}

\begin{figure}
  \centering
  \includegraphics[width=0.8\textwidth]{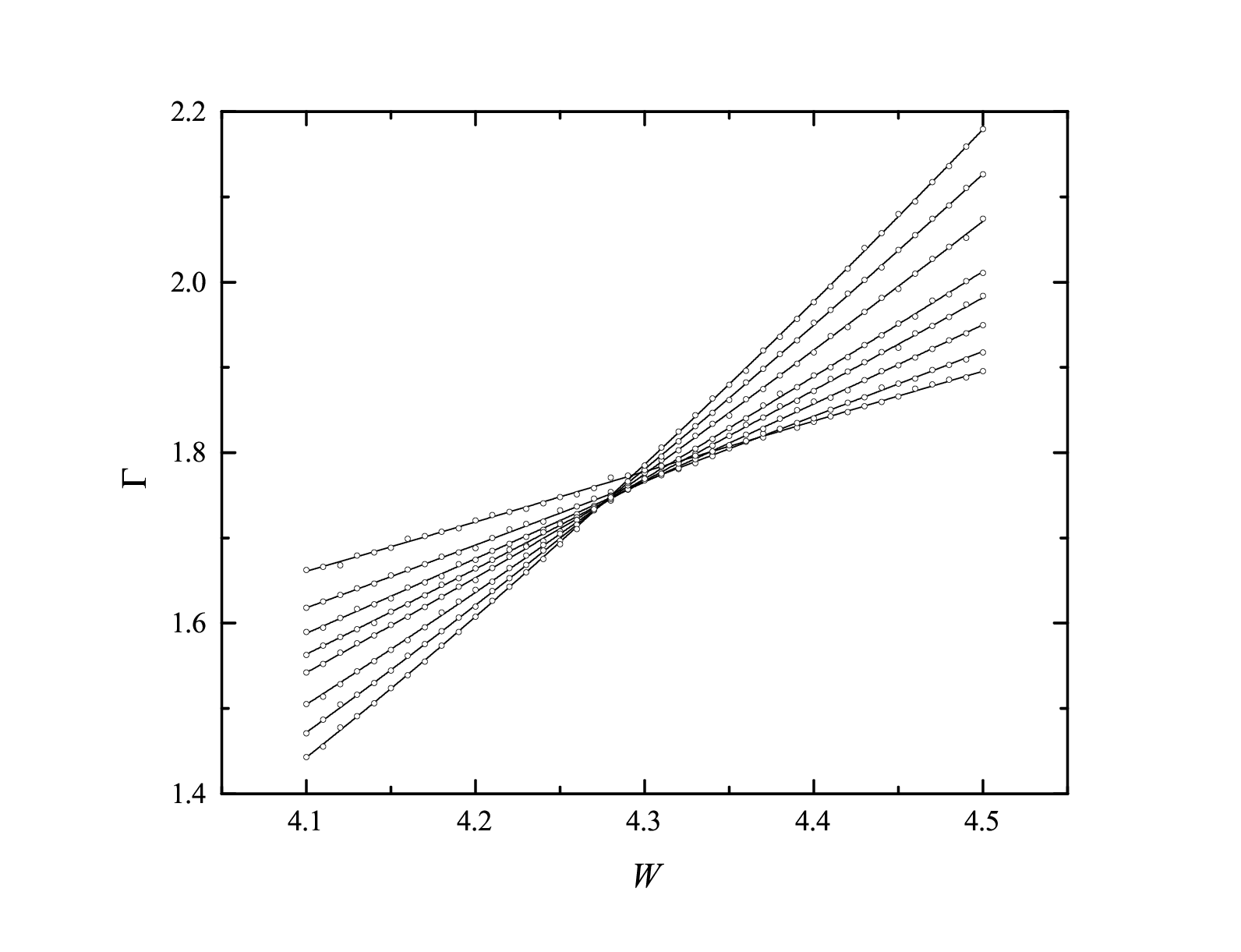}\\
  \caption{Fit of the data for the Cauchy distribution of the random potential.}\label{fig:acy}
\end{figure}

For each data set, a series of fits were performed with $m_1=1, 2$, $m_2 = 0, 1, 2$, $n_1 = 1, 2, 3, 4$ and $n_2 = 1$.
This step was automated using a combination of fortran and Python scripting.
Any fit with an unacceptable goodness of fit, i.e. $P<0.05$, was discarded.
From the remaining fits, a representative fit was chosen for each distribution of the random potential.
The orders of the expansions and the values of $\chi_\mathrm{min}^2$ are listed in Table \ref{table:details}.
The estimates of the critical disorder, $\Gamma_\mathrm{c}$, the critical exponent $\nu$, and the irrelevant exponent $y$
are listed in Table \ref{table:results}.
The data and the fit for box, normal and Cauchy distributions of the random potential are displayed in
Figures \ref{fig:abx}, \ref{fig:anm} and \ref{fig:acy}, respectively.

\begin{table}[htb]
\begin{center}
\begin{tabular}{|l|l|l|l|l|}
  \hline
   $p(W_i)$ & Disorder & $L$ & Orders of Expansions & Details of Fit \\ \hline
   box & $W\in[15,18]$ & $\ge4$ & $m_1=2, m_2=2, n_1=3, n_2=1$ & $N_{\mathrm{D}}=248, \chi_\mathrm{min}^2=239$ \\
   normal & $\sigma \in [5.75,6.55]$ & $\ge4$ & $m_1=2, m_2=1, n_1=3, n_2=1$ & $N_{\mathrm{D}}=328, \chi_\mathrm{min}^2=317$ \\
   Cauchy & $W \in [4.1,4.5]$ & $\ge4$ & $m_1=2, m_2=1, n_1=2, n_2=1$ & $N_{\mathrm{D}}=328, \chi_\mathrm{min}^2=318$ \\ \hline
   box & $W \in [15,18]$ & $\ge12$ & $m_1=2, n_1=3$ & $N_{\mathrm{D}}=124, \chi_\mathrm{min}^2=112$ \\
   normal& $\sigma \in [5.75,6.55]$ & $\ge12$ & $m_1=2, n_1=3$ & $N_{\mathrm{D}}=164, \chi_\mathrm{min}^2=158$ \\
  \hline
\end{tabular}
\end{center}
\caption{The range of data, the orders of the expansions, the total number of data,
and the value of $\chi^2_{\mathrm{min}}$ obtained in the finite size scaling analysis.
For the box distribution the energy $E=1$, and for the normal and Cauchy distributions $E=0$.}
\label{table:details}
\end{table}

\begin{table}[htb]
\begin{center}
\begin{tabular}{|l|l|l|l|l|}
  \hline
   $p(W_i)$ & $W_\mathrm{c}$ & $\Gamma_\mathrm{c}$ & $\nu$ & $y$ \\ \hline
  box & $16.536[.531,.543]$ & $1.7339[.7314,.7371]$ & $1.573[.562,.582]$ & $-3.3[-3.9,-2.8]$ \\
  normal   & $21.293[.287,.304]$ & $1.7371[.7351,.7411]$ & $1.566[.549,.576]$ & $-3.1[-4.0,-2.1]$ \\
  Cauchy  & $4.2707[.2680,.2731]$ & $1.7318[.7266,.7360]$ & $1.576[.546,.594]$ & $-2.0[-2.4,-1.7]$ \\ \hline
  box & $16.532[.526,.538]$ & $1.7316[.7286,.7345]$ & $1.577[.568,.586]$ & not applicable \\
  normal  & $21.291[.284,.298]$ & $1.7364[.7340,.7388]$ & $1.571[.560,.583]$ & not applicable \\
  \hline
\end{tabular}
\end{center}
\caption{The results of the finite size scaling analyses. Details of the simulations are given in the
corresponding row of Table \ref{table:details}.}
\label{table:results}
\end{table}

\begin{figure}
  \centering
  \includegraphics[width=0.8\textwidth]{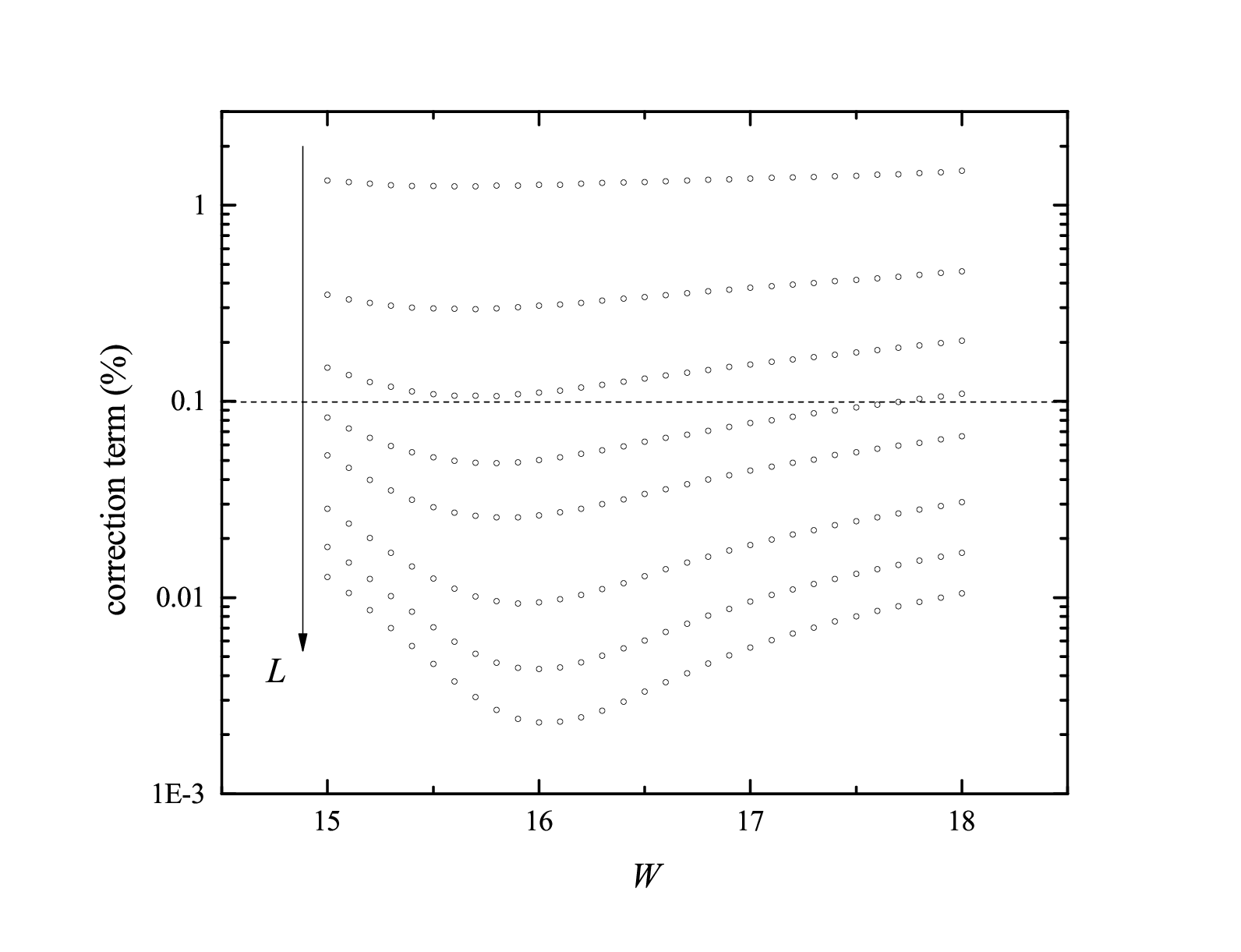}\\
  \caption{The relative value of the irrelevant correction to scaling in the fit of the data for the box distribution of the random potential.}
  \label{fig:abxcorrections}
\end{figure}

In Figure \ref{fig:abxcorrections}, we plot the contribution of the irrelevant correction for the fit to the data for
the box distribution of the random potential,
i.e. we plot the modulus of the sum of the terms with $j_2=1$ in Eq. (\ref{eq:scalingfunctionexpansion})
expressed as a percentage of the zero order term (the sum of the terms with $j_2=0$).
It can be seen that the correction is rapidly decaying with system size and that once $L>10$ the correction term is
smaller than the precision of our data.
For the normal distribution of the random potential the correction was also found to decay rapidly and become negligible
compared to the precision of our data for $L>10$.
For the Cauchy distribution of the random potential, however, the decay was much slower and the corrections are still comparable with
the precision of our data even for the largest system sizes.
Therefore, for the box and normal distributions, it  seemed reasonable to fit data for larger system sizes $L\ge 12$
without a correction due to an irrelevant scaling variable.
The details and results of these fits are also
 listed in Tables \ref{table:details} and \ref{table:results}.

\section{Discussion}

We expect that the critical exponent $\nu$ and the quantity $\Gamma_\mathrm{c}$ are universal, i.e. they depend only on the universality class
and not on other details of the model.
The results we have presented in Table \ref{table:results} are clearly consistent with this expectation
and confirm the results of our previous analysis \cite{slevin99} based on data for smaller system sizes.
In addition, while the irrelevant exponent is not very precisely determined, it seems reasonable to conclude that the irrelevant
correction for the Cauchy distribution of the random potential is different from that for the box and normal distributions.
There seem to be two possible alternative explanations for this.
The first possibility, which we think unlikely, is that there are two fixed points.
The scaling at the critical point of the box and normal distributions is controlled by one, and at the critical point of the
Cauchy distribution by the other fixed point.
Coincidentally, both these fixed points have the same (or at least very close) values of $\nu$ and $\Gamma_c$.
The second possibility, which we consider more likely, is that there is only one fixed point but that the critical points for the box and normal distributions are positioned
in the relevant space in such a way relative to the critical surface as to miss the direction associated with the smallest irrelevant index.
This would be the case, for example, if the irrelevant correction of smallest index were associated in some way with distributions which do not have second moments.

In Table \ref{table:exponents} we give the weighted average of our estimates for the critical exponent for the box, normal and Cauchy distributions
of the random potential, i.e. a weighted average of the values in the first three rows of Table \ref{table:results}.
Our result is sightly below the value $\nu=1.590[1.579,1.602]$ obtained from scaling analysis of the multi-fractal spectrum \cite{alberto10,alberto11}.
There is also a clear difference with the value $\nu=1.5$ obtained from the formula proposed by Garcia-Garcia \cite{garcia08}, which demonstrates
that his semi-classical theory is not exact.

In Table \ref{table:exponents} we also compare our result for the three dimensional orthogonal universality class with published results for the three dimensional unitary and symplectic universality classes.
The breaking of time reversal symmetry changes the exponent by roughly ten percent and
the breaking of spin-rotation symmetry changes the exponent by a slightly larger but similar amount.
The estimates of the exponents for the three dimensional symplectic and unitary classes differ only by a few percent, which is similar to the precision of the estimates themselves.
It remains a challenge to reliably distinguish exponents for these latter universality classes in a numerical simulation.

Our results for the critical exponent cannot be compared with the results of experiments on metal-insulator transitions observed in disordered electronic systems because interactions between electrons are neglected in  Anderson's model
of localisation
and how these affect the critical behaviour is not yet known.
Nevertheless, we can compare our result with the value $\nu=1.63\pm.05$ found in measurements of the dynamical localisation transition observed in a realisation of the quasi-periodically quantum kicked rotor in a cold atomic gas \cite{lopez12}.
Our numerical results are consistent with this experimental measurement.

\begin{table}[htb]
\begin{center}
\begin{tabular}{|c|c|}
\hline
$\nu$ & Universality Class \\
\hline\hline
$1.571[.563, .579]$  & 3D orthogonal (this paper)   \\
\hline
$1.43[.39, .47]$  & 3D unitary \cite{slevin97}    \\
\hline
$1.375[.359,.391]$ & 3D symplectic \cite{asada05}  \\
\hline
\end{tabular}
\caption{List of critical exponents for the three dimensional orthogonal, unitary and symplectic university classes.}
\label{table:exponents}
\end{center}
\end{table}

%%%%%%%%%%%%%%%%%%%%%%%%
Before concluding, we comment on the Anderson transition in two dimensions.
It is commonly believed that states are always localized in two dimensions \cite{aalr79}.
In fact, this is true only for the orthogonal symmetry class.
The other nine symmetry classes exhibit an Anderson transition in two dimensions.
This includes both the unitary class (Class A) and the symplectic class (Class AII).
As described below, in both cases, the finite size scaling method has played a vital role.

The transition between quantum Hall plateaux in the integer quantum Hall effect (QHE)
that occurs in the unitary class in high perpendicular magnetic fields is an Anderson transition.
The critical exponent $\nu$ has been well studied both experimentally \cite{li2009} and numerically.
For the Chalker-Coddington model \cite{chalker88,kramer05}, the exponent is estimated to be $\nu\approx 2.6$
\cite{slevin09,obuse10,amado11,fulga11,obuse12} (and see Table \ref{table:exponents2D}).
The universality of this value is supported by a study of the quantum Hall transition using a
periodically driven Hamiltonian
model \cite{dahlhaus11}.
These results, however, disagree with the experimentally measured value $\nu\approx 2.38$ \cite{li2009}.
The origin of this discrepancy has not yet been determined but there is a strong suspicion that
 it originates in the neglect of electron-electron interactions in the numerical simulations.

Systems with symplectic symmetry are realized in the presence of strong spin-orbit interaction.
Such systems have been attracting renewed interest because
the insulating phase is now known to be classified into ordinary and topological insulators
\cite{Bernevig06}.
The critical exponent $\nu$ for the transition between the metal and the ordinary insulator transition  is
estimated to be $\nu \approx 2.75$ \cite{asada02,asada04} (and see Table \ref{table:exponents2D}).
Most estimates of the exponent for the metal to topological insulator transition
\cite{obuse07,Kobayashi12,yamakage13} are consistent with the conjecture \cite{Fu12} that the exponent for both transitions is the same.
The exception is the quite different value $\nu \approx 1.6$ found \cite{Onoda07} in a numerical analysis of the metal-topological
insulator transition in the Kane-Mele model \cite{kane05}.
This discrepancy has not yet been explained.

%%%%%%%%%%%%%%%%%%%%%%%%

\begin{table}[htb]
\begin{center}
\begin{tabular}{|c|c|}
\hline
$\nu$ & Universality Class \\
\hline\hline
$2.593 [.587, .598]$ & 2D unitary (QHE)  \cite{slevin09}   \\
\hline
$2.746 [.737, .755]$  & 2D symplectic \cite{asada04}    \\
\hline
\end{tabular}
\caption{The critical exponents for the plateau transition in the integer quantum Hall effect and the 2D
symplectic university class.}
\label{table:exponents2D}
\end{center}
\end{table}

\section*{Acknowledgments}
The authors acknowledge F. Evers, A. Furusaki, H. Obuse, and L. Schweitzer for fruitful discussions.
This work was supported by  Grants-in-Aid for Scientific Research (C) (Grants No. 23540376)
and Grants-in-Aid 24000013.

\appendix
\section{}
\label{appendix}

The Wigner-Dyson classification needs to be extended to take into account  discrete symmetries,
in particular, the chiral and particle-hole symmetries, that occur in certain disordered systems\cite{gade91,gade93,zirnbauer96,altland97}.
The classification is based on Lie algebra, and in addition to the Wigner-Dyson classes, there are
3 chiral and 4 Bogoliubov de Gennes classes.
Here we summarize how the Wigner-Dyson classes discussed in this paper
are classified according to Lie algebra.

Consider an $N\times N$ Hermitian matrix $H$ and set $X=iH$. $X$
is anti-Hermitian,
is an element of the Lie algebra $\mathrm{u}(N)$, and $\exp(X)$
is an element of the Lie group $\mathrm{U}(N)$.
In the absence of any additional symmetries nothing further can be said, in general, about the Hamiltonian.
This is the unitary class in the Wigner-Dyson classification.

Any Hermitian matrix may be decomposed as $H=H_1+i H_2$, where
$H_1$ is a real symmetric matrix while $H_2$ a real antisymmetric matrix.
The matrices $H_2$ are the elements of a Lie algebra that is a subalgebra of $\mathrm{u}(N)$ with
the corresponding Lie group $\mathrm{SO}(N)$, which is a subgroup of $\mathrm{U}(N)$.
The tangent space to the symmetric space $\mathrm{U}(N)/\mathrm{O}(N)$ is the space of
real symmetric matrices  (up to a factor $i$).
The orthogonal class consists of real symmetric matrices, i.e. it spans
$\mathrm{U}(N)/\mathrm{O}(N)$.

For the symplectic class (class AII in Table \ref{tab:altland}) we must consider spin.
When spin is included in the description, the number of
degrees of freedom is doubled. The Hamiltonian is a $2N \times 2N$ Hermitian matrix,
which may be decomposed into $2\times 2$ blocks $c_{ij}$
containing matrix elements between up and down spin states.
Each block may be expressed in the form
\begin{equation}
\label{eq:quaternion1}
c_{ij}=(a_{ij}^0+ib_{ij}^0) \tau_0+(a_{ij}^1+ib_{ij}^1) \tau_1+
(a_{ij}^2+ib_{ij}^2) \tau_2+(a_{ij}^3+ib_{ij}^3) \tau_3\,,
\end{equation}
with $\tau_0=1_2$ the $2$-dimensional identity matrix, $\tau_k=i\sigma_k\,(k=1,2,3)$ with
$\sigma_k$ the Pauli matrices, and $a_{ij}^k, b_{ij}^k \, (k=0,1,2,3)$  real numbers.
Since $H$ is Hermitian,  $a_{i,j}$ and $b_{i,j}$ must satisfy
\begin{equation}
a_{ij}^0=a_{ji}^0\,,\
a_{ij}^k=-a_{ji}^k\, (k=1,2,3)\,,\
b_{ij}^0=-b_{ji}^0\,,\
b_{ij}^k=b_{ji}^k\, (k=1,2,3)\,.
\end{equation}
By use of (\ref{eq:quaternion1}), a general Hamiltonian is decomposed into $H=H_1 +H_2$
where $H_1$ is a matrix with $c_{ij}$ of the form
\begin{equation}
\label{eq:quaternion2}
c_{ij}= a_{ij}^0 \tau_0+ a_{ij}^1 \tau_1+
a_{ij}^2 \tau_2+a_{ij}^3 \tau_3\,,
\end{equation}
and $H_2$ is expressed by the $b_{ij}^k \,(k=0,1,2,3)$, i.e. the remainder.
We then define $X = i H_2$. The matrices $X$ satisfy
\begin{equation}
J X +X^\mathrm{T} J=0\,,\ J_{ij}=\delta_{ij} \tau_2\,,
\end{equation}
and are the elements of a Lie algebra that is a subalgebra of $\mathrm{u}(2N)$ with
corresponding Lie group $\mathrm{Sp}(2N)$.
The tangent space to the symmetric space $\mathrm{U}(2N)/\mathrm{Sp}(2N)$ spans the space of matrices $H_1$ (up to a factor $i$).
The Hamiltonians of systems belonging to the symplectic class are quaternion real,
and of precisely this form.
%(Incidentally, this means that the Hamiltonians of such systems can be expressed as $N \times N$ matrices %of quaternions, which can simplify both analytic and numerical calculations.)
%This is the symplectic class in the Wigner-Dyson classification.

\begin{table}[htb]
\begin{center}
\begin{tabular}{|c|c|c|c|c|c|}
\hline
WD&$[H,T]$ &$[H,S]$&Symmetric space&Symbol\\
\hline
Orthogonal&0& 0 $\forall S$& $\mathrm{U}(N)/\mathrm{O}(N)$&AI\\
\hline
Symplectic&$0$& $\neq 0$& $\mathrm{U}(2N)/\mathrm{Sp}(2N)$&AII\\
\hline
Unitary&$\neq 0$& -- & $\mathrm{U}(N)$&A\\
\hline
\end{tabular}
\end{center}
\caption{Universality classes, corresponding Lie group (symmetric spaces)
and Cartan symbol.  The first column is the Wigner-Dyson classification.
In the second and the third columns, the commutation relations of $H$ with
the time reversal operator $T$ and the spin rotation operator $S$ are shown.
The 4th column indicates the corresponding symmetric spaces, and the 5th column
the Cartan symbols for the symmetry classes of the Hamiltonian.}
\label{tab:altland}
\end{table}

Random Hamiltonians can be mapped to non-linear sigma models \cite{efetov80,hikami81}.
Reflecting the symmetries of the Hamiltonian, the non-linear sigma models are associated with
different symmetric spaces.
More details can be found in review articles such as \cite{evers08,kramer05, zirnbauer96}.

\section*{References}
\bibliography{references}

\section{Errata}

\subsection*{Caption of Figure 1}

In the original version of the article, the system size was given incorrectly as $L_y=L_z=12$. The data shown are for $L_y=L_z=10$. The caption has been corrected.

\subsection*{Table 3}

The table has been corrected to make clear that, for the normally distributed random potential, the disorder range given refers to the range of the standard deviation $\sigma$ not the parameter $W$ (see Eq. \ref{eq:normalvariance}).

\subsection*{Table 4}

For the normal distribution, the estimate of the critical value $\sigma_c$ of the standard deviation $\sigma$ was given rather than the critical value $W_c$ of the parameter $W$. The table has been corrected and now shows the estimate of $W_c$.

\subsection*{Figure 3}

The abscissa was labeled as the parameter $W$ but the data were incorrectly plotted versus the standard deviation $\sigma$. The data are now correctly plotted versus $W$.

\end{document}